\documentclass[osajnl,twocolumn,showpacs,superscriptaddress,english,10pt]{revtex4-1} 
\usepackage{amsmath,amssymb,graphicx}
\usepackage[T1]{fontenc}
\usepackage[latin9]{inputenc}
\usepackage{xargs}[2008/03/08]

\makeatletter

\newcommand{\lyxmathsym}[1]{\ifmmode\begingroup\def\b@ld{bold}
  \text{\ifx\math@version\b@ld\bfseries\fi#1}\endgroup\else#1\fi}

\usepackage{babel}

\makeatother

\usepackage{babel}
\begin{document}
\newcommandx\ket[1][usedefault, addprefix=\global, 1=]{|#1\rangle}

\newcommandx\bra[1][usedefault, addprefix=\global, 1=]{\langle#1|}

\newcommandx\avg[1][usedefault, addprefix=\global, 1=]{\langle#1\rangle}

\newcommandx\var[1][usedefault, addprefix=\global, 1=]{\langle(\Delta#1)^{2}\rangle}

\global\long\def\MM{\mathcal{M}}

\title{Photon counts statistics of squeezed and multi-mode thermal states
of light on multiplexed on-off detectors}

\begin{abstract}
Photon number resolving detectors can be highly useful for studying
the statistics of multi-photon quantum states of light. In this work
we study the counts statistics of different states of light measured
on multiplexed on-off detectors. We put special emphasis on artificial
nonclassical features of the statistics obtained. We show new ways
to derive analytical formulas for counts statistics and their moments.
Using our approach we are the first to derive statistics moments for
multi-mode thermal states measured on multiplexed on-off detectors.
We use them to determine empirical Mandel parameters and recently
proposed subbinomial parameters suitable for tests of nonclassicality
of the measured states. Additionally, we investigate subpoissonian
and superbunching properties of the two-mode squeezed state measured
on a pair of multiplexed detectors and we present results of the Fano
factor and second-order correlation function for these states. 
\end{abstract}

\author{Rados\l aw Chrapkiewicz}

\email{Corresponding author: radekch@fuw.edu.pl}

\affiliation{Institute of Experimental Physics, Faculty of Physics, University of Warsaw, ul. Ho\.z{}a 69, Warsaw, Poland}

\date{\today}


\maketitle

\section{Introduction}

Multi-photon states of light are highly applicable in precise quantum
metrology \cite{Giovannetti2004,Giovannetti2011,Demkowicz-Dobrzanski2012}.
Among these states, the N00N states \cite{Mitchell2004,Afek2010}
and the squeezed states \cite{Walls1983} are of the particular importance.
The latter type of states have recently been successfully applied,
for instance in increasing the precision of gravitational interferometer
\cite{LIGO2011} or in low-noise quantum imaging \cite{Brida2010}.

Recent advances in technology allow for measurement of photon statistics
of light with photon number resolving (PNR) detectors. They have lived
to see many implementations among which\textbf{ }the most popular
are multiplexed on-off detectors based on the photon chopping concept
\cite{Paul1996}. They have been manufactured as fiber loop detectors
\cite{Banaszek2003,Rehacek2003,Achilles2004a}, multi-pixel photon
counters (MPPC) \cite{Afek2009} or as single photon-sensitive cameras
\cite{Haderka2005,Blanchet2008}. Other types of detectors also in
use include calorimetric transition-edge sensors \cite{Brida2012}
and hybrid photo-detectors \cite{Allevi2010,Lamperti2014,Allevi2013}.
The mutltiplexed on-off detectors have clear advantages such as a
fast response and a relatively easy construction, since many of them
are based on fast avalanche photodiodes. 

Up till now there have been a number of successful experiments using
PNR detectors, thus developing further the knowledge on quantum states
of light and its sources \cite{Lamperti2014,Machulka2014,Allevi2010,Allevi2013,Bartley2013}.
Many other experiments may be enhanced by the use of PNR detectors,
for example, the observation of macro-micro entanglement \cite{Lvovsky2013}
or quantum imaging \cite{Brida2010}.

Here in this paper we will focus on the counting properties of the
multiplexed on-off detectors, which are the type of PNR detectors
most often used. These detectors alter the counts statistics as compared
to the photon statistics of light illuminating a detector \cite{Paul1996,Lee2004a,Lundeen2008,Sperling2012}.
A modification in counts statistics often leads to seemingly nonclassical
properties of measured light \cite{Bartley2013,Chrapkiewicz2014b};
therefore, an appropriate interpretation of the counts statistics
is indispensable.

In particular when it comes to an experiment, adequate criteria of
nonclassicality based on empirical counts statistics have to be applied
\cite{Sperling2012a,Kiesel2012,Sperling2013}. Criteria of nonclassicality
for counts typically allow us to determine qualitatively whether one
is dealing with a quantum state or a classical one. In order
to better differentiate between these states, the counts statistics
models are indispensable. In many cases the measured state can be
identified only based on the analysis of the mean and the variance
of counts.

In this paper we put a special emphasis on the multimode thermal states
of light for which we are the first to derive the analytical model
of counts statistics. Currently there is a great interest in the community
to observe multimode light, not only generated in the spontaneous
parametric down-conversion, but also in atomic systems including room-temperature
alkali metals vapors \cite{Boyer2008,Boyer2008b,McCormick2007,Chrapkiewicz2012}.
These experiments could be naturally reimplemented into PNR regime
with the use of single-photon sensitive cameras as in \cite{Dabrowski2014}. 

Furthermore, in this article we present the numerical model for two-mode
squeezed states observed on two identical PNR detectors. We present
the results for nonclassical measures such as the Fano factor and
the second order correlation function, commonly used in experiments.

\section{Detector counting properties}

Multiplexed detectors consist of a finite number of $N$ Geiger-type
on-off detectors. Number of detectors $N$ is the parameter which
determines the counting statistics of the detector. $N$ is also the
maximum number of counts that can be measured.

The detector operation can be described by giving the conditional
probability of obtaining $k$ counts provided that the detector was
illuminated by $n$ photons \cite{Paul1996}: 
\begin{equation}
p_{N}(k|n)=\frac{1}{N^{n}}\begin{pmatrix}N\\
k
\end{pmatrix}k!S(n,k),\label{eq:pkn}
\end{equation}
where $S(n,k)=\frac{1}{k!}\sum_{i=0}^{k}(-1)^{i}\begin{pmatrix}k\\
i
\end{pmatrix}(k-i)^{n}$ is the Stirling number of the second kind \cite{Weisstein}. For
instance, Eq. (\ref{eq:pkn}) yields $p_{N}(0|n)=0$ for all $n\geq1$
and $p(1|n)=\frac{1}{N^{n-1}}$.

\begin{figure}
\includegraphics[width=8cm]{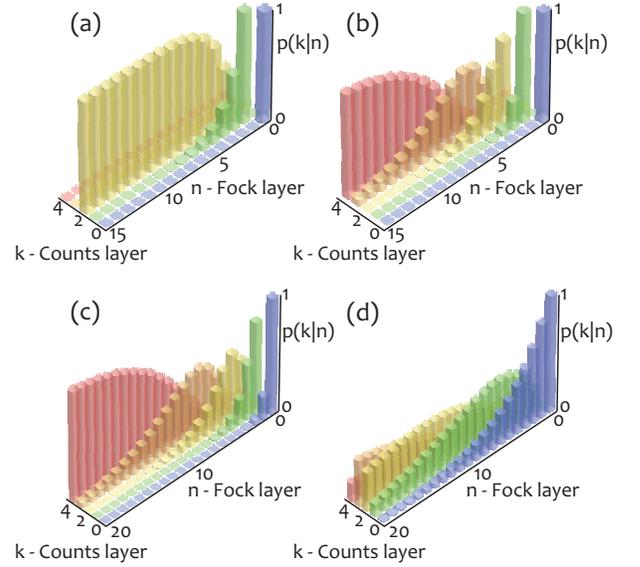}\centering

\protect\caption{Values of conditional probabilities $p_{N}(k|n)$ of getting $k$
counts when a detector is illuminated by exactly $n$ photons, for
a detector consisting of $N$ multiplexed component on-off detectors
with quantum efficiency $\eta$. (a) $N=2$, $\eta=1$ (b) (a) $N=4$,
$\eta=1$ (c) $N=4$, $\eta=0.8$ (d) $N=4$, $\eta=0.2$.\label{fig:POVMy}}
\end{figure}

Typically it is enough to know the conditional probabilities $p_{N}(k|n)$
for a finite value of $n$ since the detector saturates for large
$n$ as we illustrated in Fig. \ref{fig:POVMy} (a-b).

Due to finite quantum efficiency $\eta$ of the detector the model
Eq. \ref{eq:pkn} has to include photon losses $p_{\eta}(n|m)=\begin{pmatrix}m\\
n
\end{pmatrix}\eta^{n}(1-\eta)^{m-n}$.

This can also be done analytically for specific $n$ and $k$, for
instance $p_{N}(0|n)=(1-\eta)^{n}$ or 
\begin{equation}
p_{N}(1|n)=N(1-\eta)^{n}\left(\left(\frac{\eta-\eta N+N}{N-\eta N}\right)^{n}-1\right).
\end{equation}

We have gathered a whole set of conditional probabilities in Fig.
\ref{fig:POVMy} (c-d).

Here in this paper we derive an analytical and numerical model for
an idealized detector subject to certain assumptions. We shall assume
that the component detectors are very similar, thus principally they
will have the same quantum efficiency $\eta$. Moreover, the dark
counts and the cross talk between the component on-off detectors will
be excluded from our model. 

If one cannot apply these assumptions to the used detector, another,
alternative approach should be used such as Bayessian approach to
treat the cross-talk \cite{Kalashnikov2012,Kalashnikov2012a,Afek2009}
Including all imperfections in a general form to the theoretical model
would make it very complicated and of little use. Instead, one can
find the values of $p_{N}(k|n)$ through detector tomography \cite{Lundeen2008,Feito2009,Chrapkiewicz2014b}.

\section{Counts statistics on a single detector}

In the experiment with the PNR detector, we have access to the history
of counts collected by the detector, which yields empirical probability
distributions for counts statistics $c_{k}$. It can be expressed
as $c_{k}=\sum_{n=k}^{\infty}p_{N}(k|n)f_{n}$, where $f_{n}$ is
the photon number distribution in the measured state.

Detector losses can be taken into account both by an appropriate modification
of the conditional probabilities of the detector $p_{N}(k|n)$ and,
as described earlier, in the photons statistics.

In the following discussion we shall assume the statistics $f_{n}$
after losses and the detector will be described by no-loss conditional
probabilities as in Eq. (\ref{eq:pkn}).

In the experiment we often confine ourselves to tracking only the
mean and the variance of the counts statistics. Determination of these
values is often sufficient to identify the measured state. Now, we
shall derive the moments of probability distributions for the counts.

In general we can find average values of polynomial functions $\chi(k)$
of counts:

\begin{equation}
\avg[\chi(k)]=\sum_{k=0}^{N}\chi(k)\begin{pmatrix}N\\
k
\end{pmatrix}k!\sum_{n=k}^{\infty}\frac{1}{N^{n}}S(n,k)f_{n}\label{eq:chik}
\end{equation}

To evaluate Eq. (\ref{eq:chik}) we can use two useful properties
of Stirling number of the second kind \cite{Weisstein}: 
\begin{equation}
\sum_{n=k}^{\infty}S(n,k)\frac{x^{n}}{n!}=\frac{1}{k!}(e^{x}-1)^{k}\label{eq:Stirling1}
\end{equation}
\begin{equation}
\sum_{n=k}^{\infty}S(n,k)x^{n}=\frac{(-1)^{k}}{(1-\frac{1}{x})_{k}}\label{eq:Stirling2}
\end{equation}

where $(\cdot)_{n}$ denotes the Pochhammer symbol: $(y)_{n}\equiv\Gamma(y+n)/\Gamma(y)$.

These two properties facilitate the determination of means and variances
for states of light frequently used in experiments. For instance,
using the property Eq. (\ref{eq:Stirling1}) one can find the mean
$\avg[k]$ and the variance $\var[k]$ of counts for the coherent
state $f_{n}^{\mathrm{coh}}=\frac{\avg[n]^{n}}{n!}e^{-\avg[n]}$:

\begin{equation}
\avg[k]=N(1-e^{-\avg[n]/N})\label{eq:mean-obs-coh}
\end{equation}

\begin{equation}
\var[k]=N(1-e^{-\avg[n]/N})e^{-\avg[n]/N}\label{eq:var-obs-coh}
\end{equation}

\begin{figure}
\includegraphics[width=8cm]{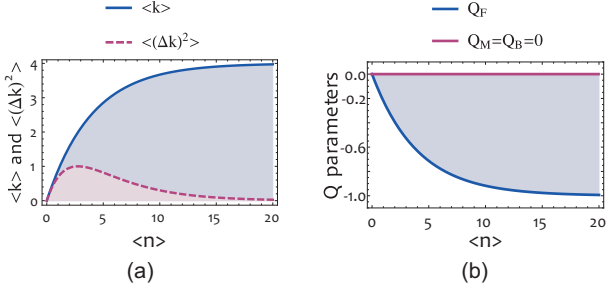}\centering

\protect\caption{ (a) Mean and variance of counts for coherent state on a detector
with $N=4$. (b) Measured $Q_{F}$ parameter is artificially nonclassical
contrary to $Q_{B}=Q_{M}=0$.\label{fig:MeanVarQM-Coh}}
\end{figure}

Here we see that the detector reduces both the mean and the variance
of counts as compared with the values for the photon statistics Fig.
\ref{fig:MeanVarQM-Coh} (a).

This reduction leads to the seemingly nonclassical properties of the
measured light. One of nonclassicality criteria for single-mode light
is the negativity of the Mandel parameter $Q_{M}=\var[n]/\avg[n]-1$.
The parameter can also be evaluated for counts statistics $Q_{F}=\var[k]/\avg[k]-1$,
here for coherent states being always negative $Q_{F}=e^{-\avg[n]/N}-1<0$
(Fig. \ref{fig:MeanVarQM-Coh} (b)).

A recently proposed modified criterion of nonclassicality is based
on the subbinomial parameter \cite{Sperling2012a,Bartley2013}:

\[
Q_{B}=N\frac{\var[k]}{\avg[k](N-\avg[k])}-1,
\]

negative only for nonclassical states and for coherent states reconstructing
the true value of $Q_{M}$ Mandel parameter (Fig. \ref{fig:MeanVarQM-Coh}
(b)).

\section{Multi-mode thermal states of light}

Now we proceed to calculations for single and multi-mode thermal states
of light which can viewed as a single subsystem of the squeezed states
of light.

At first, let us consider the following photons statistics:

\begin{equation}
f_{n}=ab^{n},\label{eq:abn}
\end{equation}

which can be readily associated with the single mode thermal state
of the mean $\avg[n]$ for which $a=(1+\avg[n])^{-1}$ and $b=\avg[n]/(1+\avg[n])$.

Then, using the property Eq. (\ref{eq:Stirling2}), we find the average
of any function of counts for given statistics $f_{n}$:

\begin{equation}
\avg[\chi(k)]_{f}=a\sum_{k=0}^{N}\chi(k)\begin{pmatrix}N\\
k
\end{pmatrix}k!\frac{(-1)^{k}}{(1-N/b)_{k}}.
\end{equation}

We focus particularly on the first and second moment of the counts
statistics:

\begin{equation}
\avg[k]_{f}=\frac{abN}{(b-1)(b(N-1)-N)}\label{eq:k1}
\end{equation}

\begin{equation}
\avg[k^{2}]_{f}=\frac{ab(b+1)N^{2}}{(b-1)(b(N-2)-N)(b+N-bN)}.\label{eq:k2}
\end{equation}

We are also able to find moments of counts distribution for another
type of statistics yielded from Eq. (\ref{eq:abn}), in particular:

\begin{equation}
g_{n}=a\frac{(n+m)!}{n!}b^{n+m},\label{eq:gn}
\end{equation}

which can be expressed as:

\begin{equation}
g_{n}=a\partial_{b,m}b^{n+m}=\partial_{b,m}b^{m}f_{n}.
\end{equation}

Note that $g_{n}$ can be readily associated with the multi-mode thermal
state with an average number of photons $\avg[n]$ and the number
of modes $\MM$ is described by the statistics $g_{n}$ Eq. (\ref{eq:gn})
for $a=\frac{1}{\Gamma(\MM)}\big(\frac{\MM}{\avg[n]+\MM}\Big)^{\MM}$,
$b=\avg[n]/(\MM+\avg[n])$ and $m=\MM-1$ \cite{Goodman2000}.

To calculate the moments for given statistics $g_{n}$ we simply do
the following:

\begin{equation}
\avg[\chi(k)]_{g}=\partial_{b,m}b^{m}\avg[\chi(k)]_{f}\label{eq:chik_gn}
\end{equation}

In this way, we can find the mean and the variance for a single-mode
thermal state, where$a=(1+\avg[n])^{-1}$ and $b=\avg[n]/(1+\avg[n])$,
similarly as in \cite{Sperling2012a}:

\[
\avg[k]_{\mathrm{Th},1}=\frac{\avg[n]N}{\avg[n]+N}
\]

\[
\var[k]_{\mathrm{Th},1}=\frac{\avg[n]N^{2}(\avg[n]N+\avg[n]+N)}{(\avg[n]+N)^{2}(2\avg[n]+N)}
\]

Equations (\ref{eq:chik_gn}), (\ref{eq:k1}) and (\ref{eq:k2}) and
substitutions for $a$ and $b$ yield analytical expressions for the
mean $\avg[k]_{\mathrm{Th},\MM}$ and the variance $\var[k]_{\mathrm{Th},\MM}$
for each $\MM$ and $N$. For example, we give analytical expressions
for $\avg[k]_{\mathrm{Th},\MM}$ for two-mode $\MM=2$ thermal state:

\[
\avg[k]_{\mathrm{Th},\MM=2}=\frac{\avg[n]N(\avg[n]+4N)}{(\avg[n]+2N)^{2}}.
\]

Further the second moment $\avg[k^{2}]_{\mathrm{Th},\MM=2}$ can be
expressed analytically:

\begin{eqnarray*}
\avg[k^{2}]_{\mathrm{Th},\MM=2} & =\\
= & \frac{\avg[n]N^{2}\left(\avg[n]^{3}+6\avg[n]^{2}N+3\avg[n]N(2N+1)+4N^{2}\right)}{(\avg[n]+N)^{2}(\avg[n]+2N)^{2}}
\end{eqnarray*}

Analytical formulas for a higher number of modes can be readily found
using any symbolic computation software and instead of presenting
directly the formulas we gather the results for higher number of modes
on plots.

\begin{figure}
\includegraphics[width=8cm]{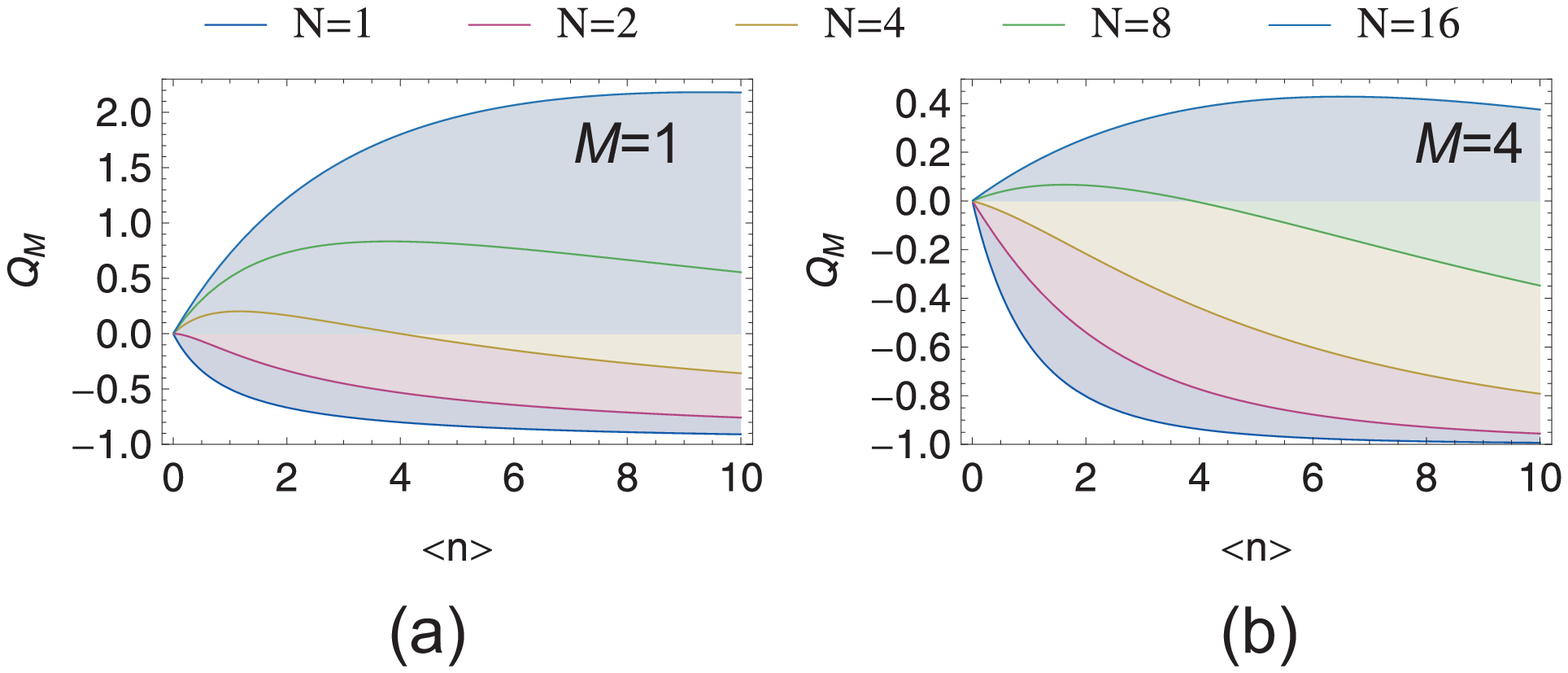}\centering

\protect\caption{Theoretical values for empirical $Q_{F}$ parameters for (a) a single-mode
thermal state, (b) multi-mode thermal state $\protect\MM=4$ plotted
versus mean number of photons $\protect\avg[n]$ for different $N$.
Depending on the parameters of the state and the detector counts statistics
may seem classical or artificially nonclassical.\label{fig:QMvsN}}
\end{figure}

Having found means and variances parameters, we can construct Mandel
parameters for counts and compare them with theoretical values $Q_{M}=\avg[n]/\MM$.
In Fig. \ref{fig:QMvsN} we show the empirical $Q_{F}$ parameter
for detectors of different $N$ for single- and multi-mode thermal
state ($\MM=4$). Depending on the mean number of input photons $\avg[n]$
and $N$, the measured $Q_{F}$ parameters appear to be nonclassical
is certain regimes.

\begin{figure}
\includegraphics[width=8cm]{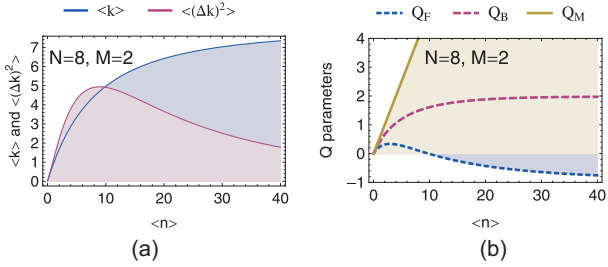}

\protect\caption{(a) Mean and variance of counts, (b) input $Q_{M}$, measured $Q_{F}$
and binomial $Q_{B}$ parameters versus mean number of photons in
two-mode ($\protect\MM=2$) thermal state on a detector with $N=8$.
\label{fig:QM-QB-QF}}
\end{figure}

\section{Counts properties of squeezed states}

The above results can be measured in a single subsystem for single-
and multi-mode squeezed states using a single detector. If we have
two detectors, we can calculate the joint counts statistics $c_{k_{1},k_{2}}$
related to the input statistics $f_{n_{1},n_{2}}$ by relation:

\[
c_{k_{1},k_{2}}=\sum_{n_{1}=k_{1}}^{\infty}\sum_{n_{2}=k_{2}}^{\infty}p_{N}(k_{1}|n_{1})p_{N}(k_{2}|n_{2})f_{n_{1},n_{2}}
\]

Calculations of the joint count statistics allow us to detect subpoissonian
correlations between the two subsystems. A good measure to quantify
these correlations is the Fano factor \cite{Lamperti2014} which,
evaluated for counts, can be expressed as:

\[
R=\frac{\var[(k_{1}-k_{2})]}{\avg[k_{1}]+\avg[k_{2}]},
\]

which is $R\geq1$ for all classical states.

Another commonly used parameter to characterize in experiments the
two modes states of light is the second order correlation function,
which evaluated on counts statistics can be expressed as:
\[
g^{(2)}=\frac{\avg[k_{1}k_{2}]}{\avg[k_{1}]\avg[k_{2}]}.
\]

It can be used to test if the investigated light demonstrate the superbunching
properties i.e. for $g^{(2)}>2$, which is characteristic for squeezed
vacuum states \cite{Iskhakov2012a} but in principle may be also obtained
using interference of thermal states \cite{Hong2012}. 

Now we shall focus on properties of joint counts statistics of the
two-mode squeezed state of light:

\[
|\psi\rangle=\sqrt{1-|\zeta|^{2}}\sum_{n=0}^{\infty}\zeta^{n}|nn\rangle,
\]

where $\zeta$ is the squeezing parameter. The Fano factor for such
a state without losses equals $R=0$, whereas the second orderd correlation
function becomes always over two: $g^{(2)}=1+1/|\zeta|^{2}\geq2.$

\begin{figure}
\includegraphics[width=8cm]{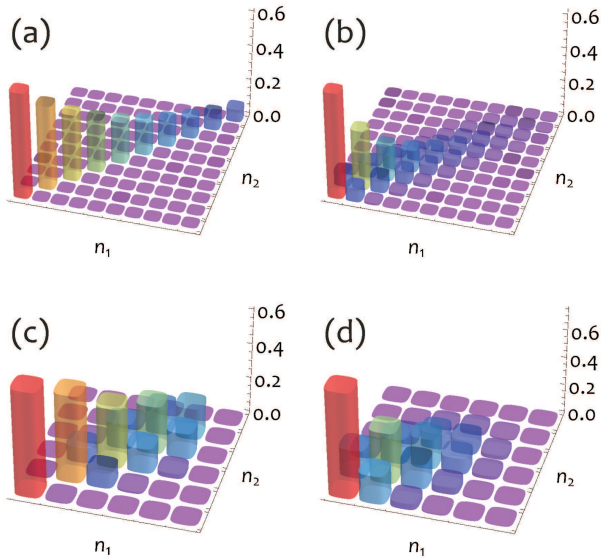}

\protect\caption{Two mode squeezed state with $\zeta=0.8$ measured by two multiplexed
detectors. (a) $N=\infty$, $\eta=1$. (b) $N=\infty$, $\eta=0.8$.
(a) $N=4$, $\eta=1$. (a) $N=4$, $\eta=0.5$.\label{fig:Sqz-examples}}
\end{figure}

In Fig. \ref{fig:Sqz-examples} we show the effect of a finite number
$N$ of component detectors and influence of losses on the joint counts
statistics $c_{k_{1},k_{2}}$, for $\zeta=0.8.$

\begin{figure}
\includegraphics[width=8cm]{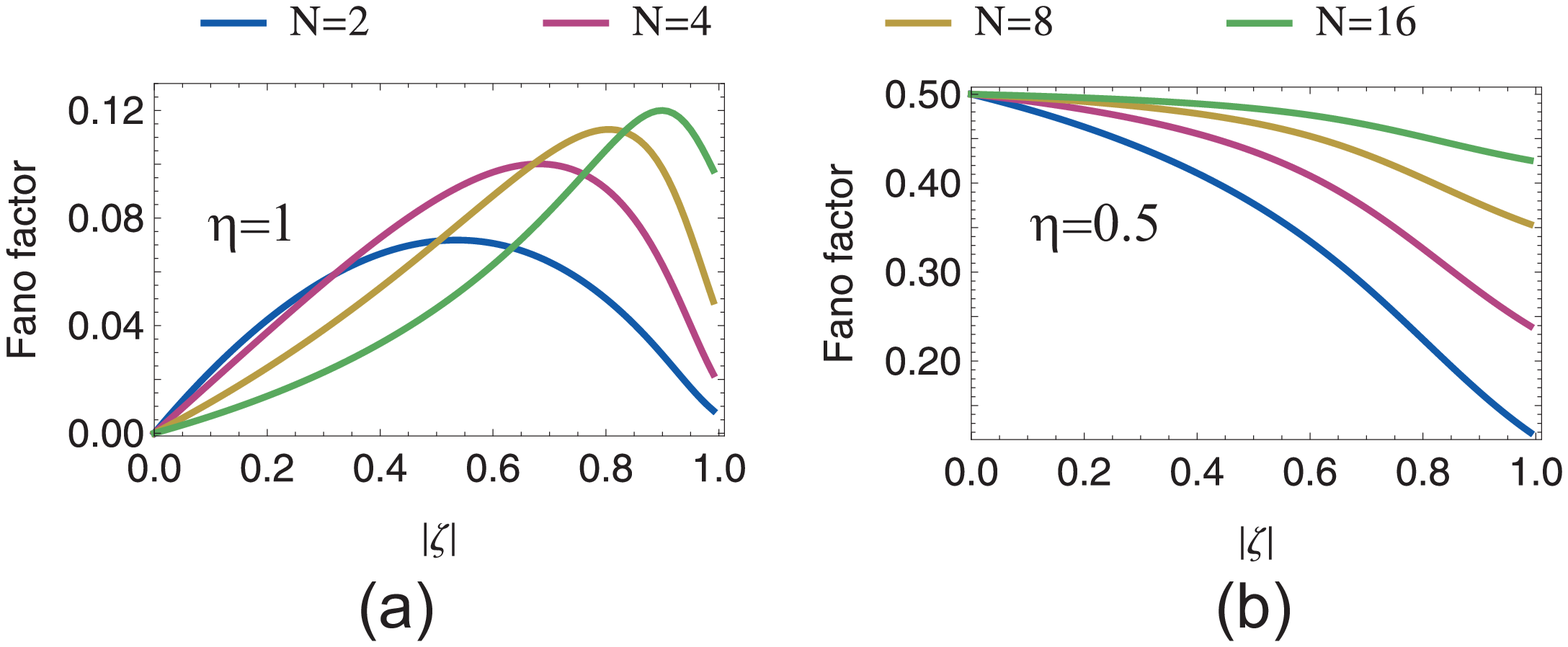}\centering

\protect\caption{Fano factors versus squeezing parameter $\zeta$ for different numbers
of $N$ detectors with (a) $\eta=1$ (b) $\eta=0.5$. A lower number
of detectors $N$ can artificially increase the correlation and consequently
decrease the Fano factor.\label{fig:FanoFactors}}
\end{figure}

Both the finite number of detectors and the losses have influence
on reducing the correlation between counts measured on two detectors.
To better understand this mechanism, we found the Fano factors for
different $N$, $\eta$ and $\zeta$. Fig. \ref{fig:FanoFactors}
(a) presents pure influence of a finite $N$, without the losses.

For small squeezing parameters $\zeta$ high number of component detectors
$N$ ensures a smaller Fano factor, whereas for high $\zeta$ the
effect is opposite. This is the effect of artificial increasing of
correlations due to the detector saturation. The saturation effects
are even more significant for the detector with finite quantum efficiency
(Fig. \ref{fig:FanoFactors} (b)), where $\eta=0.5$. Here only a
smaller number of detectors $N$ has an effect on reducing the Fano
factor for each$\zeta$.

In all these cases, the calculated counts statistics always remain
nonclassical. In order to perform a reliable test of nonclassicality
with no \emph{a priori} knowledge of the state of light, one can apply
the recently proposed criteria \cite{Sperling2013}.

\begin{figure}
\includegraphics[width=8cm]{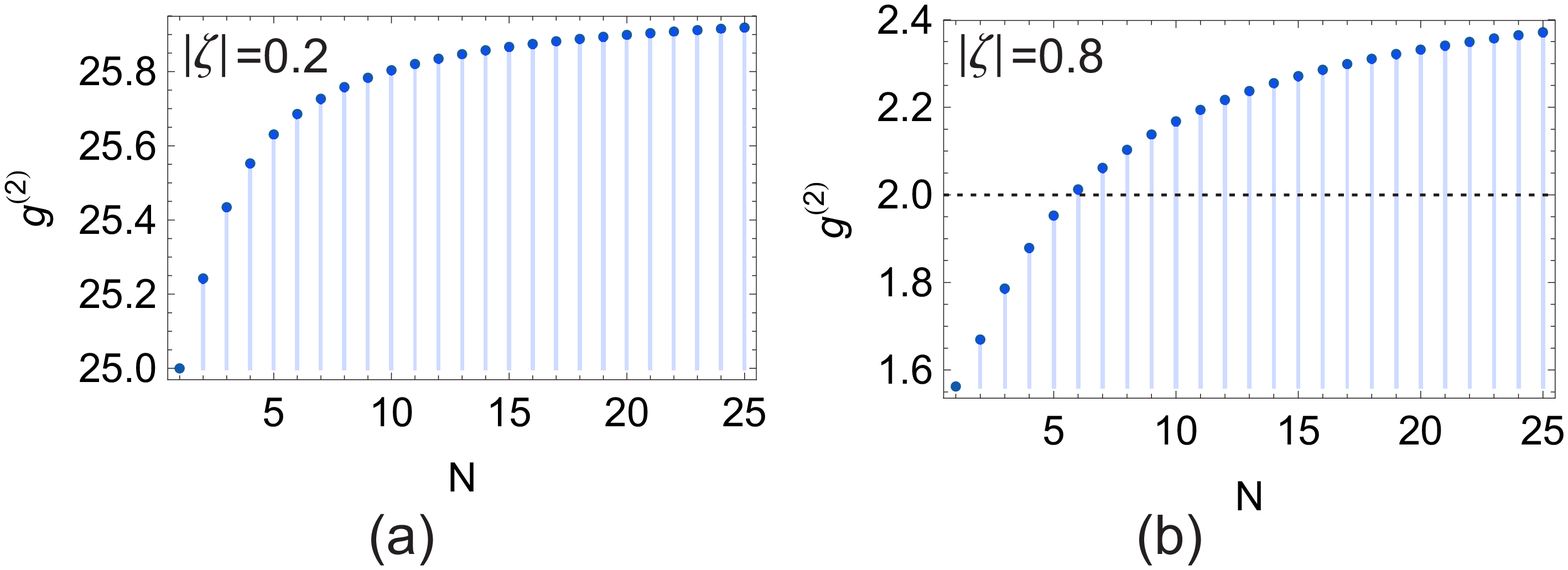}\centering

\protect\caption{Second order correlation function $g^{(2)}$ versus  different
numbers $N$ of detectors for $\eta=1$ for $|\zeta|=0.2$ (a) and
for $|\zeta|=0.8$ (b) for two-mode squeezed states. Finite number
$N$ of component on-off detectors decreases $g^{(2)}$ which for
two-mode squeezed state always is higher than two, $g^{(2)}>2$. $g^{(2)}$
is decreased by no more than one. For $|\zeta|^{2}>1/2$
and sufficiently low $N$, the observed counts statistics does not
preserve its initial superbunching properties as in (b).\label{fig:g2}}
\end{figure}

It is also instructive to view how the second correlation function
$g^{(2)}$ is modified due to the limitations introduced by a finite
number of component on-off detectors $N$. It can be seen in Fig.
\ref{fig:g2} that the low number of component on-off detectors $N$
decreases $g^{(2)}$ by no more than 1. This means that limit between
superbunched states and the bunched states ($g^{(2)}=2)$ will be
exceeded for a sufficiently low $N$ for squeezed states of the squeezing
parameter higher than $|\zeta|^{2}>1/2$ as it is exemplified in Fig.
\ref{fig:g2} (b).


\section{Summary}

In this paper, we investigate the statistics for counts measured by
a detector illuminated by squeezed or thermal states of light, the
latter being a traced subsystem of the squeezed state.

In particular, we provide a universal manner of deriving statistics
moments which we have used to determine analytical formulas for multi-mode
thermal states. In this work we put special emphasis on the issues
related to artificial nonclassicality of measured statistics in the
context of Mandel and subbinomial parameters.

On the other hand, we show the influence of increasing and decreasing subpoissonian
counts correlations for two-mode squeezed states of light measured
on two multiplexed detectors. We show that the Fano factor always
remains nonclassical. Moreover we provide the results for the second-order
correlation function $g^{(2)}$ which could drop below two in certain
cases.

The results of the work can be used to identify the states of light
measured by multiplexed on-off detectors, in particular to determine
the number of modes in thermal states. The results can significantly
contribute to designing experiments using multiplexed on-off detectors.
They will play an increasingly important role in the future since
the state of the art, photon number resolving detectors are becoming
a very efficient tool for studying the multimode, multi-photon quantum
states of light.

\section*{Acknowledgments}

I acknowledge Micha\l{} Parniak for the careful reading of the manuscript.
The project was financed by the National Science Centre grant DEC-2013/09/N/ST2/02229.


\begin{thebibliography}{10}

\bibitem{Giovannetti2004}
Vittorio Giovannetti, Seth Lloyd, and Lorenzo Maccone.
\newblock {Quantum-enhanced measurements: beating the standard quantum limit.}
\newblock {\em Science (New York, N.Y.)}, 306(5700):1330--6, 2004.

\bibitem{Giovannetti2011}
Vittorio Giovannetti, Seth Lloyd, and Lorenzo Maccone.
\newblock {Advances in quantum metrology}.
\newblock {\em Nature Photonics}, 5(4):222--229, 2011.

\bibitem{Demkowicz-Dobrzanski2012}
Rafal Demkowicz-Dobrzanski, Jan Kolodynski, and Madalin Guta.
\newblock {The elusive Heisenberg limit in quantum-enhanced metrology}.
\newblock {\em Nature Communications}, 3:1063, 2012.

\bibitem{Mitchell2004}
M~W Mitchell, J~S Lundeen, and A~M Steinberg.
\newblock {Super-resolving phase measurements with a multiphoton entangled
  state.}
\newblock {\em Nature}, 429(6988):161--4, 2004.

\bibitem{Afek2010}
Itai Afek, Oron Ambar, and Yaron Silberberg.
\newblock {High-NOON states by mixing quantum and classical light.}
\newblock {\em Science (New York, N.Y.)}, 328(5980):879--81, 2010.

\bibitem{Walls1983}
D.~F. Walls.
\newblock {Squeezed states of light}.
\newblock {\em Nature}, 306(5939):141--146, 1983.

\bibitem{LIGO2011}
{The LIGO Scientific Collaboration}.
\newblock {A gravitational wave observatory operating beyond the quantum
  shot-noise limit}.
\newblock {\em Nature Physics}, 7(12):962--965, 2011.

\bibitem{Brida2010}
G.~Brida, M.~Genovese, and I.~{Ruo Berchera}.
\newblock {Experimental realization of sub-shot-noise quantum imaging}.
\newblock {\em Nature Photon.}, 4(4):227--230, 2010.

\bibitem{Paul1996}
H.~Paul, P.~T\"{o}rm\"{a}, T.~Kiss, and I.~Jex.
\newblock {Photon Chopping: New Way to Measure the Quantum State of Light}.
\newblock {\em Physical Review Letters}, 76(14):2464--2467, 1996.

\bibitem{Banaszek2003}
Konrad Banaszek and Ian~A. Walmsley.
\newblock {Photon counting with a loop detector}.
\newblock {\em Optics Letters}, 28(1):52, 2003.

\bibitem{Rehacek2003}
J.~Rehacek, Z.~Hradil, O.~Haderka, J.~Perina, and M.~Hamar.
\newblock {Multiple-photon resolving fiber-loop detector}.
\newblock {\em Physical Review A}, 67(6):061801, 2003.

\bibitem{Achilles2004a}
Daryl Achilles, Christine Silberhorn, Cezary Sliwa, Konrad Banaszek, Ian~A.
  Walmsley, Michael~J. Fitch, Bryan~C. Jacobs, Todd~B. Pittman, and James~D.
  Franson.
\newblock {Photon-number-resolving detection using time-multiplexing}.
\newblock {\em Journal of Modern Optics}, 51(9-10):1499--1515, 2004.

\bibitem{Afek2009}
I.~Afek, A.~Natan, O.~Ambar, and Y.~Silberberg.
\newblock {Quantum state measurements using multipixel photon detectors}.
\newblock {\em Physical Review A}, 79(4):043830, 2009.

\bibitem{Haderka2005}
Ondrej Haderka, Jan Perina, and Martin Hamar.
\newblock {Direct measurement and reconstruction of nonclassical features of
  twin beams generated in spontaneous parametric down-conversion}.
\newblock {\em Physical Review A}, 71(3):4, 2005.

\bibitem{Blanchet2008}
Jean-Luc Blanchet, Fabrice Devaux, Luca Furfaro, and Eric Lantz.
\newblock {Measurement of Sub-Shot-Noise Correlations of Spatial Fluctuations
  in the Photon-Counting Regime}.
\newblock {\em Physical Review Letters}, 101(23):233604, 2008.

\bibitem{Brida2012}
Giorgio Brida, Luigi Ciavarella, Ivo~Pietro Degiovanni, Marco Genovese, Lapo
  Lolli, Maria~Griselda Mingolla, Fabrizio Piacentini, Mauro Rajteri, Emanuele
  Taralli, and Matteo G~A Paris.
\newblock {Quantum characterization of superconducting photon counters}.
\newblock {\em New Journal of Physics}, 14(8):085001,  2012.

\bibitem{Allevi2010}
Alessia Allevi, Maria Bondani, and Alessandra Andreoni.
\newblock {Photon-number correlations by photon-number resolving detectors.}
\newblock {\em Optics letters}, 35(10):1707--9, 2010.

\bibitem{Lamperti2014}
Marco Lamperti, Alessia Allevi, Maria Bondani, and Radek Machulka.
\newblock {Optimal sub-Poissonian light generation from twin beams by
  photon-number resolving detectors}.
\newblock {\em Journal of the Optical Society of America B}, 31(1):20--25,
  2014.

\bibitem{Allevi2013}
Alessia Allevi, Marco Lamperti, Maria Bondani, Jan Perina, Vaclav Michalek,
  Ondrej Haderka, and Radek Machulka.
\newblock {Characterizing the nonclassicality of mesoscopic optical twin-beam
  states}.
\newblock {\em Physical Review A}, 88(6):063807, 2013.

\bibitem{Machulka2014}
Radek Machulka, Ondrej Haderka, Jan Perina, Marco Lamperti, Alessia Allevi, and
  Maria Bondani.
\newblock {Spatial properties of twin-beam correlations at low- to
  high-intensity transition}.
\newblock ArXiv: 1405.1190, 2014.

\bibitem{Bartley2013}
Tim~J. Bartley, Gaia Donati, Xian-Min Jin, Animesh Datta, Marco Barbieri, and
  Ian~A. Walmsley.
\newblock {Direct Observation of Sub-Binomial Light}.
\newblock {\em Physical Review Letters}, 110(17):173602, 2013.

\bibitem{Lvovsky2013}
A.~I. Lvovsky, R.~Ghobadi, A.~Chandra, A.~S. Prasad, and C.~Simon.
\newblock {Observation of micro-macro entanglement of light}.
\newblock {\em Nature Physics}, 9(9):541--544, 2013.

\bibitem{Lee2004a}
Hwang Lee, Ulvi Yurtsever, Pieter Kok, George~M. Hockney, Christoph Adami,
  Samuel~L. Braunstein, and Jonathan~P. Dowling.
\newblock {Towards photostatistics from photon-number discriminating
  detectors}.
\newblock {\em Journal of Modern Optics}, 51(9-10):1517--1528, 2004.

\bibitem{Lundeen2008}
J.~S. Lundeen, A.~Feito, H.~Coldenstrodt-Ronge, K.~L. Pregnell, Ch. Silberhorn,
  T.~C. Ralph, J.~Eisert, M.~B. Plenio, and I.~A. Walmsley.
\newblock {Tomography of quantum detectors}.
\newblock {\em Nature Physics}, 5(1):27--30, 2008.

\bibitem{Sperling2012}
J.~Sperling, W.~Vogel, and G.~S. Agarwal.
\newblock {True photocounting statistics of multiple on-off detectors}.
\newblock {\em Physical Review A}, 85(2):023820, 2012.

\bibitem{Chrapkiewicz2014b}
Radoslaw Chrapkiewicz, Wojciech Wasilewski, and Konrad Banaszek.
\newblock {High-fidelity spatially resolved multiphoton counting for quantum
  imaging applications}.
\newblock {ArXiv:1405.4400, 2014.}

\bibitem{Sperling2012a}
J.~Sperling, W.~Vogel, and G.~S. Agarwal.
\newblock {Sub-Binomial Light}.
\newblock {\em Physical Review Letters}, 109(9):093601, 2012.

\bibitem{Kiesel2012}
T.~Kiesel and W.~Vogel.
\newblock {Complete nonclassicality test with a photon-number resolving
  detector}.
\newblock {\em Physical Review A}, 86:032119,  2012.



\bibitem{Sperling2013}
J.~Sperling, W.~Vogel, and G.~S. Agarwal.
\newblock {Correlation measurements with on-off detectors}.
\newblock {\em Physical Review A}, 88(4):043821,  2013.

\bibitem{Boyer2008}
Vincent Boyer, Alberto~M Marino, Raphael~C Pooser, and Paul~D Lett.
\newblock {Entangled images from four-wave mixing.}
\newblock {\em Science}, 321(5888):544--7,  2008.

\bibitem{Boyer2008b}
V~Boyer, A~Marino, and P~Lett.
\newblock {Generation of spatially broadband twin beams for quantum imaging}.
\newblock {\em Phys. Rev. Lett.}, 100(14):143601,  2008.

\bibitem{McCormick2007}
C~F McCormick, V~Boyer, E~Arimondo, and P~D Lett.
\newblock {Strong relative intensity squeezing by four-wave mixing in rubidium
  vapor.}
\newblock {\em Optics letters}, 32(2):178--80,  2007.

\bibitem{Chrapkiewicz2012}
Radoslaw Chrapkiewicz and Wojciech Wasilewski.
\newblock {Generation and delayed retrieval of spatially multimode Raman
  scattering in warm rubidium vapors}.
\newblock {\em Optics Express}, 20(28):29540, 2012.

\bibitem{Dabrowski2014}
Michal Dabrowski, Radoslaw Chrapkiewicz, and Wojciech Wasilewski.
\newblock {Hamiltonian design in readout from room-temperature Raman quantum
  memory}.
  \newblock {ArXiv:1406.6489, 2014.}

\bibitem{Weisstein}
Eric~W. Weisstein.
\newblock {Stirling Number of the Second Kind}.
\newblock {MathWorld--A Wolfram Web Resource.}

\bibitem{Kalashnikov2012}
Dmitry~A Kalashnikov, Si-Hui Tan, and Leonid~A Krivitsky.
\newblock {Crosstalk calibration of multi-pixel photon counters using coherent
  states.}
\newblock {\em Optics express}, 20(5):5044--51, 2012.

\bibitem{Kalashnikov2012a}
Dmitry~A Kalashnikov, Si-Hui Tan, Timur~Sh Iskhakov, Maria~V Chekhova, and
  Leonid~A Krivitsky.
\newblock {Measurement of two-mode squeezing with photon number resolving
  multipixel detectors.}
\newblock {\em Optics letters}, 37(14):2829--31, 2012.

\bibitem{Feito2009}
A~Feito, J~S Lundeen, H~Coldenstrodt-Ronge, J~Eisert, M~B Plenio, and I~a
  Walmsley.
\newblock {Measuring measurement: theory and practice}.
\newblock {\em New Journal of Physics}, 11(9):093038,  2009.

\bibitem{Goodman2000}
Joseph~W. Goodman.
\newblock {\em {Statistical Optics}}.
\newblock Wiley, 2000.

\bibitem{Iskhakov2012a}
T~Sh Iskhakov, A~M P\'{e}rez, K~Yu Spasibko, M~V Chekhova, and G~Leuchs.
\newblock {Superbunched bright squeezed vacuum state.}
\newblock {\em Optics letters}, 37(11):1919--21,  2012.

\bibitem{Hong2012}
Peilong Hong, Jianbin Liu, and Guoquan Zhang.
\newblock {Two-photon superbunching of thermal light via multiple two-photon
  path interference}.
\newblock {\em Physical Review A}, 86(1):013807,  2012.

\end{thebibliography}
\end{document}